\documentclass[a4paper,12pt]{article}

\usepackage[a4paper,text={17.2cm,22.8cm}]{geometry}
\usepackage{amsmath,amssymb,bm,graphicx,color}
\usepackage{extarrows}
\usepackage[utf8]{inputenc}
\usepackage{amsthm}
\usepackage{slashed}
\usepackage{tikz}
\usepackage{rotating}
\usepackage{hyperref}
\usepackage{array}
\usepackage{color}
\usepackage{multirow}
\usepackage{bbold}
\usepackage{cite}
\usepackage{pifont}
\usepackage{tabularx}

\allowdisplaybreaks
\begin{document}

\begin{titlepage}

\vspace*{-1.0cm}

\vspace{1.2cm}
\begin{center}
\large\bf
\boldmath
Dirac Masses and Mixings in the (geo)SM(EFT) and Beyond
\unboldmath
\end{center}
\vspace{0.2cm}
\begin{center}
{\large{Jim Talbert$^{a}$ and Michael Trott$^{a,b}$}}\\
\vspace{1.0cm}
{\sl
${}^a$ Niels Bohr Institute, University of Copenhagen, Blegdamsvej 17, 2100 Copenhagen, Denmark\\
${}^b$ Department of Physics (visitor), University of Notre Dame, Notre Dame, IN, 46556, USA}\\[0.5cm]
{\bf{E-mail}}: rjt89@cam.ac.uk, michael.trott@cern.ch
\end{center}

\vspace{0.5cm}
\begin{abstract}
\vspace{0.2cm}
\noindent
We report a set of exact formulae for computing Dirac masses, mixings, and CP-violation parameter(s) from 3$\times$3 Yukawa matrices $Y$
valid when $Y Y^\dagger \rightarrow U^\dagger \,Y Y^\dagger \, U$  under global $U(3)_{Q_L}$ flavour symmetry transformations $U$.
The results apply to the Standard Model Effective Field Theory (SMEFT) and its  `geometric' realization (geoSMEFT).
We thereby complete, in the Dirac flavour sector, the catalogue of geoSMEFT parameters derived at all orders in the
$\sqrt{2 \langle H^\dagger H \rangle} / \Lambda$ expansion.  The formalism is basis-independent, and can be
applied to models with decoupled ultraviolet flavour dynamics, as well as to models whose infrared dynamics
are not minimally flavour violating.  We highlight these points with explicit examples and, as a further demonstration
of the formalism's utility, we derive expressions for the renormalization group flow of quark masses, mixings, and CP-violation
at all mass dimension and perturbative loop orders in the (geo)SM(EFT) and beyond.
\end{abstract}
\vfil

\end{titlepage}


\tableofcontents
\noindent \makebox[\linewidth]{\rule{16.8cm}{.4pt}}


\section{\large{Introduction}}
\label{sec:INTRO}

The flavour sector of the Standard Model (SM) sources the bulk of its free parameters while simultaneously providing some of its richest phenomenology. These free parameters originate in the renormalizable Yukawa interactions between left- and right-chiral fermionic fields and the SU(2)$_L$ Higgs boson doublet ($H$),
\begin{equation}
\label{eq:SMYukawa}
\mathcal{L}_{SM}^Y \supset  Y^u_{pr}\, \overline{Q}_{L,p}\,\tilde{H}\,u_{R,r} + Y^d_{pr}\, \overline{Q}_{L,p}\,H\,d_{R,r} + Y^e_{pr}\, \overline{L}_{L,p}\,H\,e_{R,r} \,+ \text{h.c.} \,,
\end{equation}
whose couplings have non-trivial structure in the flavour indices $\lbrace p, r \rbrace$; here $\tilde{H}_j = \epsilon_{jk} H^{\dagger k}$. In the SM $Y$ are generically $3 \times 3$ complex matrices in flavour space. Field re-definitions allow one to reduce the number of parameters in this structure to only 13
free parameters in the (physical) fermion mass-eigenstate basis:  six quark masses, three quark mixing angles, one CP-violating Dirac phase, and finally three charged lepton masses. The quark mixing angles and CP phase appear in the Cabibbo-Kobayashi-Maskawa (CKM) mixing matrix,
which encodes all of the flavour-violation present in the SM's charged-current interactions.
The Yukawa operators of \eqref{eq:SMYukawa} are the only terms which explicitly break the global $\mathcal{G}_F \sim U(3)^5$ flavour symmetry
otherwise present in the SM \cite{Gerard:1982mm,Chivukula:1987py}.

Extracting values for the SM's mass and mixing parameters is critical to understanding its precision phenomenology,
and Beyond-the-SM (BSM) physics that may depend on (or even explain) this distinct flavour structure.
In practice this extraction is typically achieved via numerical methods for computing matrix eigenvalues (giving Dirac masses)
via diagonalizing these matrices with (bi-)unitary transformations.
A natural question to ask is: can one obtain analytic expressions for mass and mixing parameters given arbitrary forms for the complex Yukawa couplings ?
Developing such a formalism with only two fermion generations is straightforward, see Section \ref{sec:YUKAWA}.
However, for three generations standard diagonalization techniques become intractable for generic $Y^{u,d,e}$.

In this paper we employ flavour invariants to derive exact, compact formulae for the computation of Dirac mass, mixing, and CP violation parameters for fully generic $3 \times 3$ Yukawa couplings in not just the SM, but also its generalization
to the SM effective field theory (SMEFT) \cite{Buchmuller:1985jz,Grzadkowski:2010es},
\begin{equation}
\label{eq:SMEFTL}
\mathcal{L}_{SMEFT} = \mathcal{L}_{SM} + \sum_i \frac{C_i^{(d)}}{\Lambda^{d-4}} \mathcal{Q}_i^{(d)}\,.
\end{equation}
Here the sum runs over the complete basis of non-renormalizable operators $\mathcal{Q}_i^{(d)}$ composed of SM fields and invariant under the SM gauge group $\mathcal{G}_{SM} \equiv SU(3)_C \times SU(2)_L \times U(1)_Y$, at a
given mass dimension $d > 4$, with associated Wilson coefficients $C_i^{(d)}$.  Such operators are
induced generically when new physics is integrated out at a scale $\Lambda > \overline{v}_T \equiv \sqrt{2 \langle H^\dagger H \rangle}$,
with $\overline{v}_T$ the vacuum expectation value (vev) of the SM Higgs doublet.\footnote{For a comprehensive review of the SMEFT formalism, see \cite{Brivio:2017vri}.}
In particular, we will employ the geometric realization of the SMEFT, the geoSMEFT \cite{Helset:2020yio}, in our computation.  The geoSMEFT represents an all-orders reorganization of \eqref{eq:SMEFTL}, such that interactions are described on a curved manifold in scalar field space(s).  The degree of curvature depends on the ratio
$\overline{v}_T / \Lambda$, with the `flat' limit (where $\overline{v}_T / \Lambda \rightarrow 0$) corresponding to $\mathcal{L}_{SM}$.  The geoSMEFT factorizes into simple operator forms multiplying field-space connections, with the latter encoding SM theory parameters valid at all orders in $\overline{v}_T / \Lambda$.  Our results will therefore complete the geoSMEFT expressions in the Dirac flavour sector.
In addition we show that the formulae can be used to make predictions in ultraviolet scenarios when (e.g.) spontaneous flavour-symmetry breaking leads
to special textures for SM Yukawa matrices, and even in theories that introduce additional flavour violation
into the low-energy spectrum, as long as the global $U(3)_{Q_L}$  transformation properties of $Y Y^\dagger$ are respected.

Besides their obvious predictive utility, our formulae may also be of use in high(er)-order global SMEFT fits to existing data, especially to CKM mixing elements (see \cite{Descotes-Genon:2018foz}), or in studying non-standard flavour effects in the SMEFT (see e.g. \cite{Aebischer:2020lsx,Bruggisser:2021duo}).  Towards the latter end, we use our formulae to rapidly derive their renormalization group flow at all mass dimension and loop orders  in the (geo)SM(EFT), including an explicit numerical calculation of quark sector RGE at one-loop perturbative order in minimally flavour-violating (MFV) \cite{DAmbrosio:2002vsn} theories.  Our results therefore constitute a generic formalism for studying flavour in (B)SM matching cases of the SMEFT, and also open the door for related studies in the leptonic sector, when non-zero neutrino masses are properly accounted for.

The paper develops as follows:  in Section \ref{sec:YUKAWA} we review the Yukawa sector of the geoSMEFT, while in Section \ref{sec:FORMULAE} we present the unique flavour invariants we employ, and use them to derive the final formulae.  In addition, we discuss the domain of applicability of these expressions with demonstrated examples.  In Section \ref{sec:RGE} we derive generic, analytic expressions for the renormalization group flow of the flavour parameters.  Finally, we conclude in Section \ref{sec:CONCLUDE}, providing an outlook for the extension of this formalism into the lepton sector.  Some useful formulae are presented in Appendix \ref{app:RGEFORM}.

\section{\large{The Yukawa Sector of the (geo)SM(EFT)}}
\label{sec:YUKAWA}
The geoSMEFT \cite{Helset:2020yio} represents a re-organization of the SMEFT operator product expansion (OPE) in \eqref{eq:SMEFTL}, such that
\begin{equation}
\label{eq:geoREORG}
\mathcal{L}_{\text{SMEFT}} = \sum_i G_i \left(I,A, \phi, ... \right) \, f_i \,.
\end{equation}
Here $G_i$ are field-space connections that depend on the group indices $I, A$ of all internal symmetry groups, and real scalar field coordinates $\phi_I$ of the SU(2)$_L$ Higgs doublet, normalized as follows:\footnote{$\phi_4$ is expanded around the vev via $\phi_4 \rightarrow \phi_4 + \overline{v}_T$.}
\begin{equation}
\label{eq:higgscoord}
H \left( \phi_I \right) = \frac{1}{\sqrt{2}} \left[
\begin{array}{c}
\phi_2 + i \phi_1 \\
\phi_4 - i \phi_3
\end{array}
\right] \,\,\, \Longrightarrow \,\,\,
\tilde{H} \left( \phi_I \right) = \frac{1}{\sqrt{2}} \left[
\begin{array}{c}
\phi_4 + i \phi_3 \\
-\phi_2 + i \phi_1
\end{array}
\right]\, .
\end{equation}
On the other hand, $f_i$ are operator forms composed of the Lorentz-index-carrying building blocks of $\mathcal{L}_{\text{SMEFT}}$, and which are (largely)\footnote{Except powers of $D^\mu H$, which get grouped with $f_i$.  This residual scalar coordinate-dependence in the composite operator forms manifests in powers of $ \partial_\mu h$, upon the Higgs acquiring its vev.} independent of $\phi$.  That is, upon the Higgs obtaining its vev, the tower of interactions composing $G_i$ reduces to a number and emissions of $h$ (the propagating Higgs field), while $f_i$ remains a distinct operator of SM fields and derivatives.  Critically, by using Hilbert series techniques one can show that, at least for two- and three-point functions, the number of $f_i$ saturates to a constant value at arbitrary mass-dimension.  This then allows the field-space connections $G_i$, and thereby important theory parameters (e.g. gauge boson masses, gauge couplings, weak mixing angles, the Higgs mass, etc.), to be defined at all orders in the SMEFT's characteristic $\overline{v}_T / \Lambda$ expansion.  For a complete description of the geoSMEFT see \cite{Helset:2020yio}, which builds on prior work \cite{Helset:2018fgq,Corbett:2019cwl,Alonso:2015fsp,Alonso:2016oah}.  See \cite{Hays:2020scx,Corbett:2021eux,Corbett:2021jox,Corbett:2021cil} for recent geoSMEFT applications.

In what follows we are only concerned with the Yukawa sector of the effective theory, i.e. with non-renormalizable SMEFT operators of the form
\begin{equation}
\label{eq:HFops}
Q^{6+2n}_{\overset{\psi H}{\scriptscriptstyle{pr}}} = \left(H^\dagger H \right)^{n+1}\left(\bar{\psi}_{L,p} \psi_{R,r} H \right) \,\,\,\,\,\text{with} \,\,\,\,\, n \ge 0\,,
\end{equation}
with $\psi_L \in \lbrace Q, L \rbrace_L$ and $\psi_R \in \lbrace u, d, e \rbrace_R$, and where $H \rightarrow \tilde{H}$ in the second bracket when $\psi_R = u_R$. Hence the classification of two- and three-point functions in terms of their field space connections and composite operator forms, as already present in \cite{Helset:2020yio}, is sufficient for our purposes.  From there we recall the two-point Yukawa function $Y(\phi) \overline{\psi}_1 \psi_2$, whose field space connection $Y(\phi)$ is given by
\begin{equation}
\label{eq:fieldspacegen}
Y_{pr}^{\psi_1} \left(\phi_I \right) = \frac{\delta \mathcal{L}_{\text{SMEFT}}}{\delta (\overline{\psi}^I_{2,p} \psi_{1,r})} \bigg|_{\mathcal{L}(\alpha, \beta, ...) \rightarrow 0} \,,
\end{equation}
for a generic fermion sector $\psi \in \lbrace u, d, e \rbrace$.  Here $\alpha, \beta, ...$ denote effective gauge couplings etc., such that the notation $\mathcal{L}(\alpha, \beta, ...) \rightarrow 0$ implies that all Lagrangian terms and spin connections with  non-trivial Lorentz indices are sent to zero.   As a demonstrative example, the Higgs potential is given simply by $V(\phi) = - \mathcal{L}_{\text{SMEFT}} \vert_{\mathcal{L}(\alpha, \beta, ...) \rightarrow 0}$.

Using \eqref{eq:fieldspacegen} one quickly arrives at the explicit expressions for the fermionic field space connections,
\begin{align}
\label{eq:explicitfieldspace}
Y_{pr}^{\psi} \left(\phi_I\right) &= - H \left( \phi_I \right) \left[Y_\psi\right]_{pr}^\dagger + H \left( \phi_I \right) \sum_{n=0}^{\infty} C^{(6+2n)}_{\overset{\psi H}{\scriptscriptstyle{pr}}} \left(\frac{\phi^2}{2} \right)^n \,,
\end{align}
where again $H \rightarrow \tilde{H}$ when $\psi = u$, from which one defines the all-orders Yukawa couplings
\begin{equation}
\label{eq:YukAO}
[ \mathcal{Y}^\psi ]_{rp} = \frac{\delta (Y^\psi_{pr})^\dagger}{\delta h} \bigg|_{\phi_i \rightarrow 0} =  \frac{\sqrt{h}^{44}}{\sqrt{2}} \left( [Y_\psi]_{rp} - \sum_{n=3}^\infty \frac{2n-3}{2^{{n-2}}} \tilde{C}^{(2n),\star}_{\overset{\psi H}{\scriptscriptstyle{pr}}} \right)\, .
\end{equation}
Here $\sqrt{h}^{44}$ is the 44 component of the matrix square-root of the expectation value for the scalar field space metric, $\langle h^{IJ} \rangle^{1/2}$; $h^{IJ}(\phi)$ is the all-orders connection for derivative Higgs operators:  $\mathcal{L} \supset h_{IJ} (\phi) \left(D_\mu \phi \right)^I \left( D^\mu \phi \right)^J$.   Also note that the Wilson coefficients have been normalized in \eqref{eq:YukAO} such that $\tilde{C}_i^{(2n)} \equiv C_i ^{(2n)}\, \overline{v}_T^{2n-4} / \Lambda^{2n-4}$.
Finally, we recall that the fermion mass matrices are given after the Higgs acquires it vev,
\begin{equation}
\label{eq:massAO}
\left[M_\psi \right]_{rp} = \langle (Y_{pr}^\psi)^\dagger \rangle \,.
\end{equation}
The flavour structure of the theory can be analyzed in either phase, and so in what follows we generally work with the Yukawa matrices in \eqref{eq:YukAO}.
\subsection{Extension to Fermionic Mixing}
\label{sec:MIX}
Our goal is to extend the definition of all-orders parameters in the geoSMEFT to the fermion mass and mixing sector, i.e. those quantities defined in the fermion mass-eigenstate basis, where the different generations of a given fermion family (e.g. $u$, $c$, $t$) can be physically distinguished by their distinct masses.  As in the SM, this requires the diagonalization of \eqref{eq:YukAO}, which can be achieved through a bi-unitary transformation on both the SU(2)$_L$ doublet and singlet fermion fields:
\begin{equation}
\label{eq:biunitary}
[U_{\psi L}^{\dagger}]_{ir}\,[\mathcal{Y}^\psi]_{rp} \, [U_{\psi R}]_{pj} \equiv [D_\psi]_{ij} = \text{diag} \left(y_{\psi 1}, y_{\psi 2}, y_{\psi 3} \right)\,.
\end{equation}
One can rotate away the dependence on the singlet transformations by constructing the Hermitian combinations $\mathcal{Y} \mathcal{Y}^\dagger$ and diagonalizing these objects through unitary field transformations, i.e.
\begin{equation}
\label{eq:mmdag}
[U_{\psi L}^{\dagger}]_{ir}\,[\mathcal{Y}^\psi \mathcal{Y}^{\psi,\dagger}]_{rp} \, [U_{\psi L}]_{pj} = \text{diag} \left(y^2_{\psi 1}, y^2_{\psi 2}, y^2_{\psi 3} \right)\,.
\end{equation}
Dropping the $L$ subscript from the mixing matrices, one then writes the physical CKM matrix as the overlap between up and down-quark fermion mixing:
\begin{equation}
\label{eq:CKMdef}
V_{CKM} \equiv U^\dagger_{u} U_d \equiv  \left(
\begin{array}{ccc}
V_{ud} & V_{us} & V_{ub} \\
V_{cd} & V_{cs} & V_{cb} \\
V_{td} & V_{ts} & V_{tb} \\
\end{array}
\right)\,.
\end{equation}
As mentioned in the introduction, $V_{CKM}$ can be generically parameterized in terms of three real mixing angles $\theta_{ij}$ and a Dirac CP-violating phase $\delta$, and we do so with the standard Particle Data Group (PDG) \cite{Zyla:2020zbs} representation,
\begin{equation}
\label{eq:CKMgen}
V_{CKM}  = \left(
\begin{array}{ccc}
c_{12} c_{13} & s_{12} c_{13} & s_{13}\, e^{-i \delta}\\
-s_{12} c_{23} - c_{12} s_{23} s_{13}\, e^{i \delta} & c_{12} c_{23} - s_{12} s_{23} s_{13}\, e^{i \delta} & s_{23} c_{13} \\
s_{12} s_{23} - c_{12} c_{23} s_{13}\, e^{i \delta} & -c_{12} s_{23} - s_{12} c_{23} s_{13}\, e^{i \delta} & c_{23} c_{13}\\
\end{array}
\right)\,,
\end{equation}
where $\cos \theta_{ij} \equiv c_{ij}$, etc.  Note that while the structural form of this matrix is identical to its SM counterpart, the mixing angles encode the fact that there is an infinite tower of higher-order operator forms contributing to the Yukawa connection (and therefore flavour violation), as shown in \eqref{eq:YukAO}.  For notational simplicity we do not differentiate these `effective' mixing elements from those in the flat limit where $C \rightarrow 0$.

As a final comment, we remind the reader that, while we choose to focus on the all-orders geoSMEFT, which resolves various ambiguities relating to field redefinitions and operator bases in the effective theory (amongst other benefits), our conclusions hold at any given fixed order in the `standard' SMEFT as well.

\section{\large{All-Orders Masses, Mixings, and CP-Violation}}
\label{sec:FORMULAE}
Our goal in this section is to derive expressions for the Yukawa eigenvalues $y_i^2$ and CKM mixing parameters $\theta_{ij}$ and $\delta$, in terms of $\mathcal{Y}_{rp}$ alone, such that these Lagrangian parameters are defined at all orders in the $\overline{v}_T/\Lambda$ expansion characterizing the geoSMEFT.  To this end we first consider the two-generation case, where one can expand the flavour indices of $\mathcal{Y}$, finding
\begin{equation}
\label{eq:YukAO2D}
\mathcal{Y} = \frac{\sqrt{h}^{44}}{\sqrt{2}} \left[  \left(
\begin{array}{cc}
Y_{11} & Y_{22} \\
Y_{21} & Y_{22}
\end{array}
\right) - \sum_{n=3}^\infty \frac{2n-3}{2^{{n-2}}} \left(
\begin{array}{cc}
\tilde{C}^{(2n),\star}_{11} & \tilde{C}^{(2n),\star}_{21} \\
\tilde{C}^{(2n),\star}_{12} & \tilde{C}^{(2n),\star}_{22}
\end{array}
\right) \right]\,.
\end{equation}
Such an expansion holds for both quark families $\mathcal{Y}_{u,d}$.  To avoid bi-unitary transformations, we next form the Hermitian product $\mathcal{Y} \mathcal{Y}^\dagger$, and take its magnitude (anticipating the fact that no CP-violation is present in two-generation theories):
\begin{equation}
\label{eq:2DYuk}
\vert \mathcal{Y} \mathcal{Y}^\dagger \vert  \equiv \left(
\begin{array}{cc}
\vert \mathbb{y}_{11} \vert & \vert \mathbb{y}_{12} \vert \\
\vert \mathbb{y}_{12} \vert & \vert \mathbb{y}_{22} \vert
\end{array}
\right)\,\,\,\,\, \Longrightarrow U = \left(
\begin{array}{cc}
\cos \theta&  \sin \theta \\
-\sin \theta & \cos \theta
\end{array}
\right)\,.
\end{equation}
The off-diagonal elements of $\vert \mathcal{Y} \mathcal{Y}^\dagger \vert$ are equivalent due to Hermiticity, and one recognizes that this object is now a symmetric, real matrix, which is trivially diagonalized by $U$, leading to Yukawa eigenvalues and a mixing angle that can be expressed as
\begin{equation}
\label{eq:2Devalues}
y_{i,j}^2 = \frac{1}{2} \left( \mathbb{y}_{11} + \mathbb{y}_{22} \mp \sqrt{\mathbb{y}_{11}^2 + 4 \mathbb{y}_{12}^2 - 2 \mathbb{y}_{11}\mathbb{y}_{22} + \mathbb{y}_{22}^2}  \right)\,,\,\,\,\,\,\,\,\,\,\,\,t_{2\theta} = \frac{2 \, \vert \mathbb{y}_{12} \vert}{\left(  \vert \mathbb{y}_{22} \vert - \vert \mathbb{y}_{11} \vert \right)}\,.
\end{equation}
As $\theta$ lives in the first quadrant by definition, $\sin \theta \ge 0 $, $\cos \theta \ge 0$, and hence $\tan \theta \ge 0$.  Also, while we have derived \eqref{eq:2Devalues} in the arbitrary weak-interaction basis such that it holds in both the up and down-sectors, one can rotate without loss of generality to a flavour basis where either $Y_d$ or $Y_u$ is diagonal, respectively leading to either $U_{d}= \mathbb{1}$ or $U_{u}= \mathbb{1}$.  In this case $\theta \rightarrow \theta_C$, where $\theta_C$ is the renowned (physical) Cabibbo angle \cite{Cabibbo:1963yz} describing the dominant CKM mixing between first and second quark generations.  While \eqref{eq:2Devalues} is valid at all-orders in the geoSMEFT expansion in $\overline{v}_T/\Lambda$, there is clearly a basis dependence to the expressions which is undesirable when comparing with experiment.  Furthermore, when three fermion generations are present, obtaining analogous expressions for $y_i^2$, $\theta_{ij}$, and $\delta$ becomes intractable if approached with standard diagonalization techniques as above, regardless of concerns over the basis-dependence of the results.
\subsection{Approach with Flavour Invariants}
\label{sec:INVARIANTCALC}
To circumvent these issues we will instead employ the (rephasing and weak-basis) invariant theory developed in \cite{Jenkins:2007ip,Jenkins:2009dy,Feldmann:2015nia,Greenberg:1985mr,Dunietz:1985uy,Jarlskog:1985ht,Nieves:1987pp,Nieves:2001fc,Kusenko:1993wu,Kusenko:1993ph,Davidson:2003yk,Branco:2004hu,Branco:2005jr,Dreiner:2007yz}.  While we leave the details of the invariant theory to these prior references, we recall that flavour invariants are objects that do not change under field redefinitions corresponding to unitary transformations of SM fermions under the global $\mathcal{G}_F \sim U(3)^5$ flavour symmetry.   Such transformations correspond to basis changes of the Yukawa connections,\footnote{We focus only on the quark sector from this point forward.   Additional discussion regarding extensions to leptons will be given in Section \ref{sec:CONCLUDE}.}
\begin{equation}
\label{eq:GFgroupaction}
Y^u\, \longrightarrow \, U^\dagger_{Q_L} \, Y^u \, U_{u_R}\,, \,\,\,\,\,\,\,\,\,\, Y^d \, \longrightarrow \, U^\dagger_{Q_L} \, Y^d \, U_{d_R}\,,
\end{equation}
which, as argued above, are the fundamental objects encoding the mass and mixing parameters of the theory. Polynomials of $Y^{\psi}$ define a group ring $\mathbb{C}$ when the linear combinations of all possible products of the generators $Y^{\psi}$ (with complex coefficients) are formed. The $\mathcal{G}_F$-invariant ring  $\mathbb{C}^{\mathcal{G}_F}$ is that set of polynomials unchanged under the action $\mathcal{G}_F$ (cf. \eqref{eq:GFgroupaction}), and it can be shown that $\mathbb{C}^{\mathcal{G}_F}$ is finitely generated.  A central result of \cite{Jenkins:2009dy} is an explicit representation of said generators, denoted $I_i$ (and henceforth referred to simply as `invariants'), for the quark sector of the SM with both two and three fermion generations.  Here we extend this analysis to the geoSMEFT, and further use the invariants to extract our desired formulae.

In the case of three fermion generations, Hilbert Series techniques lead one to conclude that there are 11 polynomially independent invariants $I_i$ \cite{Jenkins:2009dy}. In what follows we will actually use the representation and notation given in \cite{Feldmann:2015nia} to present $I_i$.  Here, after defining the Hermitian combination $Y Y^\dagger \equiv \mathbb{Y}$, a complete set of 3D quark-sector invariants are reported as
\begin{align}
\nonumber
I_1 &\equiv \text{tr} \left(\mathbb{Y}_u \right)
\,, \,\,\,\,\,\,\,\,\,\,\,\,\,\,\,
\hat{I}_3 \equiv \text{tr} \left( \text{adj}\, \mathbb{Y}_u \right)
\,, \,\,\,\,\,\,\,\,\,\,\,\,\,\,\,
\hat{I}_6 \equiv \text{tr} \left(\mathbb{Y}_u \, \text{adj} \, \mathbb{Y}_u \right) = 3 \, \text{det} \,\mathbb{Y}_u \,, \\
\label{eq:massinvariants}
 I_2 &\equiv \text{tr} \left(\mathbb{Y}_d \right)
\,, \,\,\,\,\,\,\,\,\,\,\,\,\,\,\,
\hat{I}_4 \equiv \text{tr} \left( \text{adj}\, \mathbb{Y}_d \right)
\,, \,\,\,\,\,\,\,\,\,\,\,\,\,\,\,
\hat{I}_8 \equiv \text{tr} \left(\mathbb{Y}_d \, \text{adj} \, \mathbb{Y}_d \right) = 3 \, \text{det} \,\mathbb{Y}_d\,,
\end{align}
for invariants `unmixed' between up and down sectors, and
\begin{align}
 \label{eq:CKMinvariants}
 \hat{I}_5 &\equiv \text{tr} \left(\mathbb{Y}_u \, \mathbb{Y}_d \right)
 , \,\,\,\,\,\,\,\,\hat{I}_7 \equiv \text{tr} \left(\text{adj} \, \mathbb{Y}_u \, \mathbb{Y}_d \right)  ,\,\,\,\,\,\,\,\,\, \hat{I}_9 &\equiv \text{tr} \left(\mathbb{Y}_u \, \text{adj} \, \mathbb{Y}_d \right)   ,\,\,\,\,\,\,\,\,\, \hat{I}_{10} \equiv \text{tr} \left(\text{adj} \,\mathbb{Y}_u \, \text{adj} \, \mathbb{Y}_d \right) \,,
\end{align}
for real, mixed invariants. Note that it is natural to consider $\mathbb{Y}_u$ and $\mathbb{Y}_d$  when considering flavour symmetry violating effects
as these two invariants are not simultaneously diagonalizable. This is why in meson mixing, for example, flavour violation is always proportional
to such invariants.  One notes that \eqref{eq:massinvariants} contains all of the information required to extract the Yukawa eigenvalues.  On the other hand, \eqref{eq:CKMinvariants} know about the overlap between up and down sectors, and are therefore functions of the mixing elements $U^{u,d}_{ij}$.  As there are four invariants, this is sufficient to extract the four independent elements of the (unitary) CKM mixing matrix.  The expressions in \eqref{eq:massinvariants}-\eqref{eq:CKMinvariants} are all CP-even and $\ge 0$.  Additionally there is the complex, CP-Odd invariant
\begin{equation}
\label{eq:phaseinvariants}
I_{11}^- = - \frac{3 i}{8} \text{det} \left[ \mathbb{Y}_u, \mathbb{Y}_d \right] \,,
\end{equation}
which is proportional to the Jarlskog determinant \cite{Jarlskog:1985ht}.  This invariant will therefore be necessary for completely determining the sign of the Dirac CP-violating phase below.

Note that, in the above expressions,
 $\text{adj} \, \mathbb{Y}$ are the adjoint matrices\footnote{These matrices are adjoint under the $SU(3)_{Q_L}$ quark flavour subgroup of the global $U(3)^5$ SM flavour symmetry, i.e. $U(3)^5 \supset U(3)^3 / U(1)_B \sim SU(3)_{Q_L} \times U(3)_{u_R} \times U(3)_{d_R}$, with $U(1)_B$ global baryon number.} satisfying $\mathbb{Y} \, \text{adj} \,\mathbb{Y} = \text{det} \,\mathbb{Y}$, i.e.
\begin{align}
\text{adj} \, \mathbb{Y} &= \mathbb{Y}^2 - \text{tr}\left[\mathbb{Y}\right] \mathbb{Y} + \frac{1}{2} \left(\text{tr}^2\left[\mathbb{Y}\right] - \text{tr}\left[\mathbb{Y}^2 \right] \right) \mathbb{1}\,,
\end{align}
which holds for matrices that are diagonalized via unitary field transformations.  Generalized expressions for adjoint matrices diagonalized by bi-unitary transformations can also be found.  Both are derived directly from the Cayley-Hamilton identity for $3\times3$ matrices \cite{Feldmann:2015nia}.
\subsubsection*{Invariants}
A core observation of this work is that the structure of the flavour invariants is the same regardless of the mass dimension considered in the (geo)SM(EFT).  This can be seen clearly in \eqref{eq:YukAO}, where one notes that the addition of higher-order SMEFT operators simply results in an all-orders reparameterization of each individual matrix element of the Hermitian objects $\mathbb{Y}$, which compose the polynomial invariants.  Hence we can freely form \eqref{eq:massinvariants}-\eqref{eq:phaseinvariants} with explicit dependence on the geoSMEFT's Hermitian Yukawa coupling,
\begin{equation}
\label{eq:SMEFTYuk}
\mathbb{Y}_{rp} =   \frac{\mathbb{h}}{2} \, \left( Y_{ri} Y^\star_{pi} - \sum_{n^\prime}^\infty f(n^\prime) Y_{ri} \tilde{C}_{ip}^{(2n^\prime)}  - \sum_{n}^\infty f(n) \tilde{C}_{ir}^{(2n),\star} Y^\star_{pi} + \sum_{n,n^\prime}^\infty f(n) f(n^\prime) \tilde{C}^{(2n),\star}_{ir} \tilde{C}_{ip}^{(2n^\prime)}  \right),
\end{equation}
where $\mathbb{h} \equiv (\sqrt{h}^{44})(\sqrt{h}^{44})^\star$, $f(n) = \frac{2n-3}{2^{n-2}}$, and analogously for $f(n^\prime)$.  We observe that \eqref{eq:SMEFTYuk} represents the fundamental BSM object in this formalism.
\subsubsection*{Quark Masses}
Proceeding to the extraction of the quark Yukawa eigenvalues $y^2$, we solve the system of equations in \eqref{eq:massinvariants}, finding
\begin{align}
\label{eq:upquark1}
y^2_i &= \frac{\left(-2\right)^{1/3}}{3 \, \psi_u} \left(I_1^2 - 3 \, \hat{I}_3 + \left(-2\right)^{-1/3}\,I_1 \, \psi_u + \left(-2\right)^{-2/3} \, \psi_u^{2}\right)\,, \\
\nonumber
y^2_{j,k} &= \frac{1}{12 \psi_u} (\left(-2\right)^{4/3}\,I_1^2 - 3 \cdot \left(-2\right)^{4/3}\,\hat{I}_3 + 4\, I_1 \, \psi_u \\
\label{eq:upquark23}
&\mp    \psi_u \,\sqrt{24 \left(I_1^2 - 3\, \hat{I}_3 \right) + \frac{6 \cdot \left(-2\right)^{5/3} \left(I_1^2 -3\, \hat{I}_3 \right)^2}{\psi_u^2} -3 \cdot \left(-2\right)^{4/3} \psi _u^2} + \left(-2\right)^{2/3} \psi_u^2)
\end{align}
for the up-quark Yukawa eigenvalues, where the $\psi_u$ parameter is given by
\begin{equation}
\psi_u = \left(-2\, I_1^3 + 9\, I_1 \hat{I}_3 - 9\, \hat{I}_6 +
 3\, \sqrt{-3\, I_1^2 \hat{I}_3^2 + 12\, \hat{I}_3^3 + 4\, I_1^3 \hat{I}_6 - 18\, I_1 \hat{I}_3 \hat{I}_6 + 9\, \hat{I}_6^2} \right)^{1/3} \,.
\end{equation}
Recall that by definition the functions $y^2$ and $I_{1,3,6}$ are $\ge 0$, and of course $y^2$ is also real.  While certain individual expressions on the RHS of these equations are imaginary, we have checked that the eigenvalues are in fact real.
 Note also that the system of equations in \eqref{eq:massinvariants} permits six different solutions, corresponding to the six different possible mass hierarchies for $y_{u,c,t}$.  The unmixed mass invariants remain constant under the transposition of any two flavour-eigenstate indices, e.g. $I_1 = I_1 (i \leftrightarrow j)$. However, by definition, one recognizes the up quark as the lightest generation, the top as the heaviest, and the charm as the intermediate,
 \begin{equation}
 y_u^2 \equiv \text{min} \lbrace y_i^2, y_j^2, y_k^2 \rbrace \,, \,\,\,\,\,\,\,\,\,\,  y_c^2 \equiv \text{mid} \lbrace y_i^2, y_j^2, y_k^2 \rbrace \, \,\,\,\,\,\,\,\,\,\, y_t^2 \equiv \text{max} \lbrace y_i^2, y_j^2, y_k^2 \rbrace \,.
  \end{equation}
Furthermore, the expressions for the down-quark mass-eigenvalues are analogous to \eqref{eq:upquark1}-\eqref{eq:upquark23}, replacing
\begin{equation}
\label{eq:downreplace}
\lbrace I_1, \hat{I}_3, \hat{I}_6 \rbrace \longrightarrow \lbrace I_2, \hat{I}_4, \hat{I}_8 \rbrace
\end{equation}
which, when implemented, results also in the notation change $\psi_u \longrightarrow \psi_d$ in the final equations.  The beauty of \eqref{eq:upquark1}-\eqref{eq:upquark23} is that their RHS can be calculated in any arbitrary flavour basis, whilst the LHS are always the mass eigenvalues (up to the Higgs vev).
\subsubsection*{CKM Parameters}
Given the Yukawa/mass eigenvalues at all orders, one can then move to the four real CKM parameters, which can be obtained by solving the system of equations implied by $\hat{I}_{5,7,9,10,11}$.  We will solve \eqref{eq:CKMinvariants} for $\vert V_{cd} \vert^2$, $\vert V_{cs} \vert^2$, $\vert V_{td} \vert^2$, and $\vert V_{ts} \vert^2$, and then use the unitarity constraints of the CKM matrix to uniquely determine its remaining matrix elements in terms of the invariants $\hat{I}_{5,7,9,10,11}$.  We then have all of the information required to determine the three mixing angles $s_{ij}$, which upon using the final $I^{-}_{11}$ invariant will give us the phase $\delta$.  Critically, we observe that \eqref{eq:CKMinvariants} does not permit multiple solutions for $\vert V_{ij} \vert^2$.

Proceeding along these lines, it is straightforward to derive the following compact expressions for the mixing angles $s_{ij}$:
\begin{align}
\label{eq:s13}
s_{13} &= \left[\frac{-\hat{I}_{10} - y_b^2 \, \left(\hat{I}_7 - \Delta_{ds}^+ \Delta_{uc}^+ \Delta_{ut}^+\right) - y_u^2 \left(\hat{I}_9 + y_b^2 \left( \hat{I}_5 - y_b^2 \Delta_{ct}^+ \right) - y_d^2 \,y_s^2\, \Delta_{ct}^+ \right) }{\Delta_{bd}^- \Delta_{bs}^- \Delta_{cu}^- \Delta_{ut}^-}\right]^{1/2} \,, \\
\label{eq:s23}
s_{23} &= \left[\frac{\Delta_{tu}^- \left(-\hat{I}_{10} +y_c^2\left(-\hat{I}_9 + \left(y_b^4 + y_d^2 y_s^2 \right)\Delta_{ut}^+\right) + y_b^2 \left(-\hat{I}_7 + y_c^2 \left(-\hat{I}_5 + \Delta_{ct}^+ \Delta_{ds}^+ \right) + y_u^2 \Delta_{ct}^+ \Delta_{ds}^+\right) \right)}{\Delta_{ct}^- \left(\hat{I}_{10} +  y_u^2\,\hat{I}_9 + y_b^2 \left(\hat{I}_7 + y_u^2 \left(\hat{I}_5 -2 \Delta_{ct}^+ \Delta_{ds}^+ \right) \right) - \left(y_u^4 + y_c^2 y_t^2 \right) \left(y_b^4 + y_d^2 y_s^2 \right) \right)} \right]^{1/2} \\
\label{eq:s12}
s_{12} &= \left[\frac{\Delta_{db}^- \left(\hat{I}_{10} + y_s^2 \left(\hat{I}_7 - y_c^2 y_t^2 \Delta_{db}^+ \right) \right) + y_u^2\,\Delta_{bd}^- \left(-\hat{I}_9 - y_s^2\,\hat{I}_5 + \Delta_{sb}^+ \Delta_{ct}^+ \Delta_{ds}^+ \right) + y_u^4 y_s^2 \left(y_b^4 - y_d^4 \right)}{\Delta_{ds}^- \left(\hat{I}_{10} + y_u^2\, \hat{I}_9 + y_b^2 \left(\hat{I}_7 + y_u^2\left(\hat{I}_5 - 2 \Delta_{ct}^+\Delta_{ds}^+ \right)\right) -\left(y_b^4 +y_d^2 y_s^2 \right) \left(y_u^4 + y_c^2 y_t^2 \right)  \right)} \right]^{1/2} \,,
\end{align}
which are given in terms of the Yukawa eigenvalues $y_i^2$ and the difference/sum parameters $\Delta^\pm_{ij}$ defined by
\begin{equation}
\label{eq:psiparameter}
\Delta^\pm_{ij} \equiv y_i^2 \pm y_j^2.
\end{equation}
  Given the expressions in \eqref{eq:upquark1}-\eqref{eq:upquark23} for $y_i^2$, one then has the desired final expressions entirely in terms of the flavour-invariants ${I}_i$.  Note that, by definition, these angles lie in the first quadrant, $0 \le s_{ij} \le \frac{\pi}{2}$.  Finally, one can use $I^-_{11}$, along with \eqref{eq:s13}-\eqref{eq:s12}, to derive that
\begin{align}
\label{eq:delta}
s_\delta &= \frac{4}{3}\, I_{11}^- \,\left[\, \Delta_{tc}^- \Delta_{tu}^-\Delta_{cu}^-\Delta_{bs}^-\Delta_{bd}^-\Delta_{sd}^-\,s_{12} s_{13} s_{23} \left(1-s_{23}^2 \right)^{1/2} \left(1-s_{12}^2 \right)^{1/2} \left(1-s_{13}^2\right)\right]^{-1} \,.
\end{align}
As $I^-_{11}$ is not guaranteed to be $\ge 0$, one sees that \eqref{eq:delta} uniquely pins down the sign of $\delta$.  These expressions therefore complete the list of quark flavour parameters in the (geo)SM(EFT); \eqref{eq:upquark1}-\eqref{eq:delta} and their renormalization group flow (cf. \eqref{eq:y23run}-\eqref{eq:deltarun} in Section \ref{sec:RGE}) represent the core results of this work.
\subsection{Numerical Checks}
\label{sec:NUMERICS}
In order to probe the reliability and accuracy of \eqref{eq:upquark1}-\eqref{eq:delta} we performed a series of numerical tests.  Specifically we have built an automated script that\footnote{Note that we do this computation in the arbitrary flavour / weak-eigenstate basis where all matrix elements are a priori non-zero and complex.  To obtain our numerical results we use {\tt{Mathematica}}'s built-in {\tt{Eigenvalues}} and {\tt{Eigenvectors}} functional calls, and at no point in this numerical procedure was any information about the flavour invariants $I_i$ given to the extraction code.}
\begin{enumerate}
\item assembles $\mathcal{Y}^\psi_{rp}$ by randomly drawing up a list of $n - 1$, 3$\times$3 arrays of complex numbers and assigning these arrays to $Y^\psi_{rp}$ and $\tilde{C}^{(2n)}_{\psi H}$, thereby forming the effective Yukawa coupling in \eqref{eq:YukAO}. The Hermitian combination $\left[\mathcal{Y}\mathcal{Y}^\dagger \right]_{rp}$ is then constructed.
\item computes the invariants using \eqref{eq:massinvariants}-\eqref{eq:CKMinvariants}, the Yukawa eigenvalues from \eqref{eq:upquark1}-\eqref{eq:upquark23}, the mixing angles from \eqref{eq:s13}-\eqref{eq:s12}, and finally the Dirac phase from \eqref{eq:delta}.
\item simultaneously computes the eigenvalues of the Hermitian coupling obtained in {\bf{(1)}} numerically.
\item simultaneously computes the diagonalizing matrices $U_{(u,d)}$  numerically and extracts the four physical parameters of the CKM matrix. This is achieved by
\begin{enumerate}
\item computing the eigenvectors of $\left[\mathcal{Y}^{(u,d)}\mathcal{Y}^{(u,d),\dagger} \right]$.  These are normalized to unit vectors $\mathbb{v}_i$ and then the numerical matrices are defined by $U_{(u,d)} \equiv \left(\mathbb{v}_1^{(u,d), T}, \mathbb{v}_2^{(u,d), T}, \mathbb{v}_3^{(u,d), T} \right) $.
\item computing the CKM matrix as $V_{CKM} = U_u^\dagger \cdot U_d$.
\item uniquely extracting the $s_{13}$ mixing angle from $V_{13}$, $s_{13} = \vert V_{13} \vert$.
\item uniquely extracting the $s_{23}$ mixing angle from $V_{23}$, $s_{23} = \vert V_{23} \vert / \sqrt{1-s_{13}^2}$.
\item uniquely extracting the $s_{12}$ mixing angle from $V_{12}$, $s_{12} = \vert V_{12} \vert / \sqrt{1-s_{13}^2}$.
\item uniquely extracting the $s_\delta$ phase in a phase-convention-independent manner from the Jarlskog invariant $J$.
\end{enumerate}
\item subtracts the values obtained in {\bf{(4)}} from those in {\bf{(2)}}.  Their differences are added to a list, and the process is repeated, at various $n$.
\end{enumerate}
The result is that, in no draw and with no parameter, and at arbitrary mass-dimension $n$ and new physics scales $\Lambda$,\footnote{We have run scans up to $n=10$ and allowed for new physics between 2-10 TeV.} are the differences between the values computed with \eqref{eq:upquark1}-\eqref{eq:delta} in step {\bf{(2)}} and those computed with the numerical procedure between steps {\bf{(3)-(4)}} greater than the numerical tolerance we allow for ($10^{10}$ in this case).  As is clear, these checks therefore validate the expressions in \eqref{eq:upquark1}-\eqref{eq:delta} in a highly non-trivial manner.
\subsection{Applicability to other BSM Scenarios}
\label{sec:BSM}
Given \eqref{eq:upquark1}-\eqref{eq:delta}, it is interesting to note when they do (and do not) apply. Indeed, the formalism is in general complete when \eqref{eq:massinvariants}-\eqref{eq:phaseinvariants} are sufficient to extract all of the flavour parameters of the low-energy theory.  As described above, these are determined by analyzing the global flavour group ring $\mathbb{C}^{\mathcal{G}_F}$, based on the transformation properties of $Y^\psi$ in \eqref{eq:GFgroupaction}.  Hence  \eqref{eq:upquark1}-\eqref{eq:delta} hold in any theory where \eqref{eq:GFgroupaction} represents the generic transformation properties of the Dirac Yukawa/mass matrices under $\mathcal{G}_F$, such that $Y^\psi Y^{\psi\,\dagger} \rightarrow U^\dagger \, Y^\psi Y^{\psi\,\dagger}  \, U$ under global flavour transformations $U \in U(3)_{Q_L}$, including theories where $Y^\psi$ or $M^\psi$ originate from ultraviolet dynamics \cite{Jenkins:2009dy}.  Of course, if $Y^\psi Y^{\psi\,\dagger}$ respects these transformation properties, but there are additional sources of flavour violation in the infrared spectrum, \eqref{eq:massinvariants}-\eqref{eq:phaseinvariants} will be incomplete as there are just enough invariants to uniquely determine the 10 (13) flavour parameters of the quark (quark $+$ charged lepton) Yukawa sector.  More invariants would be required to extract the values of parameters associated to the additional flavour violation.

For academic purposes we also consider the case when global flavour rotations are not family-universal, i.e. when
\begin{equation}
\label{eq:Ytransform1}
\lbrace \mathbb{Y}_u, \mathbb{Y}_d \rbrace \rightarrow \lbrace \mathbb{Y}_{u}^{\prime} \, , \, \mathbb{Y}_{d}^{\prime} \rbrace =  \lbrace U^{u\,\dagger}_\chi \,\mathbb{Y}_u \, U^{u}_\chi, U^{d\,\dagger}_\chi \,\mathbb{Y}_d \, U^{d}_\chi \rbrace \,\,\,\,\,\text{with}\,\,\,\, U^{d}_\chi \neq U^{u}_\chi \,.
\end{equation}
Such a situation may be conceivable in the broken electroweak phase, e.g.  Here we observe that \eqref{eq:massinvariants} (which we denote $\hat{I}_{\text{mass}}$), and therefore \eqref{eq:upquark1}-\eqref{eq:upquark23}, still hold as basis-independent expressions, since $\hat{I}_{\text{mass}} \rightarrow  \hat{I}^\prime_{\text{mass}} = \hat{I}_{\text{mass}}$ due to the cyclic property of the Trace.  However, \eqref{eq:CKMinvariants}-\eqref{eq:phaseinvariants} do not exhibit this invariance generically.  This is clear, for example, in $\hat{I}_5$:
\begin{equation}
\label{eq:Ytransform2}
\hat{I}^\prime_5 = \text{tr} \left(\mathbb{Y}_u^\prime \mathbb{Y}_d^\prime \right) = \text{tr} \left(U_\chi^{u\,\dagger} \,\mathbb{Y}_u\,U_\chi^{u} U_\chi^{d\,\dagger} \,\mathbb{Y}_d \, U_\chi^{d} \right)  \neq \hat{I}_5 \,.
\end{equation}
This implies that predictions for CKM mixing parameters in such theories will not hold.  We have checked that these statements are true by generating random unitary transformations $U_\chi^\psi$ numerically and constructing \eqref{eq:Ytransform1}, where $y_i^2$, $\theta_{ij}$ and $\delta$ are also generated randomly.  We compare the values computed with \eqref{eq:upquark1}-\eqref{eq:delta} to those extracted numerically, and indeed confirm that while the Yukawa/mass eigenvalues are in perfect agreement with the randomly drawn inputs, the mixing parameters are not.  On the other hand, we now give two examples from the literature where our formulae apply exactly.\footnote{Note that we do not present these models in the context of a matching calculation to the (geo)SMEFT, but rather as example UV constructions with BSM field and symmetry content remaining in the spectrum --- our goal is to show that our model-independent formulae from above can be used to study the Dirac flavour sector of specific models in the literature, and as a result we only give relevant details to that end below.}
\subsubsection*{An effective theory of flavour in the ultraviolet}
Our formalism can be used to analytically or numerically compute the mass, mixing, and CP-violating parameters predicted from typical high-energy EFTs of flavour (e.g. Froggatt-Nielsen models \cite{Froggatt:1978nt}), where some ultraviolet dynamics at a scale $\Lambda_F$ breaks a flavour symmetry $\mathcal{G}_F$, yielding specifically textured infrared Yukawa couplings.  As an example we consider the quark sector of the non-Abelian `universal texture zero' (UTZ) model of \cite{deMedeirosVarzielas:2017sdv}.  Here the high-energy Lagrangian is given as an OPE of non-renormalizable interactions between SM fields and heavy scalars $\lbrace \theta_i, \Sigma, S \rbrace$,
\begin{equation}
\label{eq:UTZLag}
\mathcal{L}_{\text{UTZ}}  \supset \psi_p\left ({1\over M_{3,f}^2}\theta_3^p  \theta_3^r  +{1\over M_{23,f}^3}\theta_{23}^p\theta_{23}^r\Sigma+{1\over M_{123,f}^3}(\theta_{123}^p\theta_{23}^r+\theta_{23}^p\theta_{123}^r)  S    \right)\psi_r^{c}H + \mathcal{O}(1/M^4) + \text{...}
\end{equation}
where the labels indicate a specific family sector, $f \in \lbrace u, d, e, \nu \rbrace$, and
whose terms are invariant under the SM gauge symmetries and $\mathcal{G}_F \simeq \Delta \left(27\right)$ due to the triplet (anti-triplet) $\Delta \left(27\right)$ charge assignments of the SM multiplets (flavons).\footnote{The Higgs field is a trivial flavour singlet, as are the BSM $\Sigma$ and $S$ fields, with the latter helpful for shaping the couplings and the former associated to Grand Unified symmetry breaking.  See \cite{deMedeirosVarzielas:2017sdv} for further details of the model's implementation, including its scalar potential.}  Upon $\lbrace \theta \rbrace$ developing vevs in specific directions of flavour-space (and with specific scales $\lbrace v \rbrace $), the UTZ Lagrangian in \eqref{eq:UTZLag} populates Dirac mass matrices of the form
\begin{equation}
\label{eq:UTZmass}
\mathcal{M}^D_f = \left(
\begin{array}{ccc}
0 & a \, e^{i\gamma} & a \, e^{i\gamma}  \\
a \, e^{i\gamma}  & \left(b\,e^{-i \gamma} + 2a\,e^{-i \delta} \right) e^{i(\gamma+\delta)} & b \, e^{i\delta}  \\
a \, e^{i\gamma}  & b \, e^{i\delta}  & 1-2a \,e^{i \gamma} + b \, e^{i \delta}
\end{array} \right)_f \, ,
\end{equation}
for all Dirac fermions ($u$, $d$, ...), where the free parameters $\lbrace a, b, \gamma, \delta \rbrace_f$ are associated to ultraviolet dynamics, and where we have removed two unphysical phases according to the discussion in \cite{Roberts:2001zy}. This object is then the only information required to form the invariants in \eqref{eq:massinvariants}-\eqref{eq:CKMinvariants}, and therefore also the mass and mixing formulae in \eqref{eq:upquark1}-\eqref{eq:delta}, which yield the physical flavour parameters.  We focus on the quark sector of the model, and input the best-fit values for the unconstrained ultraviolet inputs $\lbrace a, b, \gamma, \delta \rbrace_{u,d}$ from \cite{deMedeirosVarzielas:2017sdv} and compute, finding
\begin{equation}
\frac{m_u}{m_t} = 7.16 \cdot 10^{-6}\, , \,\,\,\,\,\,\,\,\,\, \frac{m_c}{m_t} = 0.0027 \, , \,\,\,\,\,\,\,\,\,\, \frac{m_d}{m_b} = 0.00090  \, , \,\,\,\,\,\,\,\,\,\, \frac{m_s}{m_b} = 0.020 \,,
\end{equation}
for the mass eigenvalues (note that the Higgs vev is already encoded in the numerical values for the ultraviolet parameters) and
\begin{equation}
s_{12} = 0.226, \,\,\,\,\,\, s_{23} = 0.0191, \,\,\,\,\, s_{13} = 0.0042, \,\,\,\,\, s_{\delta} = 0.5609,
\end{equation}
for the mixing angles and Dirac phase, in perfect agreement with the numerical values extracted in \cite{deMedeirosVarzielas:2017sdv}, up to $\mathcal{O}(1/M_f^4)$.  Extending the analysis to higher-order operators in \eqref{eq:UTZLag} is straightforward and no more complicated.
\subsubsection*{On models with light leptoquarks}
As examples of theories with non-minimal flavour violation in the infrared, we consider models incorporating leptoquarks, exotic scalars that couple SM SU(2)$_L$ doublet quark and lepton fields, e.g.
\begin{equation}
\mathcal{L} \supset y_{pr}^{LL}\, \overline{Q}_{L,p}^{C}\, \Theta_{3}\, L_{L,r} + z_{pr}^{LL}\, \overline{Q}_{L,p}^{C}  \,\Theta_{3}^{\dagger} \, Q_{L,r} + \text{h.c.} \,,
\end{equation}
where we have given illustrative operators for a scalar triplet leptoquark $\Theta_{3}$ which transforms as $\Theta_3 \sim  \left({\bf{\bar{3}}}, {\bf{3}}, 1/3\right)$ under $\mathcal{G}_{SM}$, and where all gauge/internal index contractions are implied.\footnote{For a somewhat comprehensive review of leptoquark models and basic phenomenology, see \cite{Dorsner:2016wpm}.}  Here the couplings $y^{LL}$ and $z^{LL}$ are also $3\times3$ matrices in flavour space, and are in general not simultaneously diagonalized along with $Y^\psi$.  There will therefore be additional (physical) mixing parameters associated to the leptoquark couplings that \eqref{eq:upquark1}-\eqref{eq:delta} cannot account for. Regardless, as discussed above, \eqref{eq:upquark1}-\eqref{eq:delta} can still predict the Dirac parameters $y_i^2$, $\theta_{ij}$, and $\delta$.

To demonstrate this we focus on the toy leptoquark models developed in \cite{Bernigaud:2019bfy,Bernigaud:2020wvn}.  Here the generic representation for the Yukawa matrices in the quark sector can be written as
\begin{equation}
\label{eq:leptobasis}
Y_u^{\prime \prime} =  P^\dagger \, \Lambda_d \,V_{CKM}^\dagger \cdot \left(
\begin{array}{ccc}
y_u & 0 & 0 \\
0 & y_c & 0 \\
0 & 0 & y_t
\end{array} \right) \cdot \Lambda_U^\dagger \, P, \,\,\,\,\,\,\,\, Y_d^{\prime \prime} = P^\dagger \, \Lambda_d \,\cdot \left(
\begin{array}{ccc}
y_d & 0 & 0 \\
0 & y_s & 0 \\
0 & 0 & y_b
\end{array} \right) \cdot \Lambda_D^\dagger \, P\,,
\end{equation}
where the double prime ($\prime \prime$) notation simply indicates that, as is clear, these matrices are in a special flavour basis that is neither the traditional weak-eigenstate nor fermion mass-eigenstate basis.  That is, information about the structure of $y^{LL}$ in this model is already encoded in \eqref{eq:leptobasis} ($z^{LL}$ was not considered), through the introduction of the rotations $\Lambda_{d,U,D}$ ($P$ is associated to the representation theory of various flavour groups --- see \cite{Bernigaud:2020wvn} for details).

Hence \eqref{eq:leptobasis} are ideal candidates to test our conclusions from above, and to do so we consider one of the explicit models in \cite{Bernigaud:2020wvn}, based on a $D_{15}$ dihedral flavour symmetry.  Without presenting the details of its ultraviolet Lagrangian, we simply note that this model yields predictions for special Yukawa textures,
\begin{equation}
\label{eq:D15matrices}
Y^{\prime \prime}_u = v_u \, \left(
\begin{array}{ccc}
a_u/v_u & 0 & 0 \\
0 & b_u \cos \frac{\pi}{15} & -c_u \sin \frac{\pi}{15} \\
0 & b_u \sin \frac{\pi}{15} & c_u \cos \frac{\pi}{15}
\end{array}
\right)\,, \,\,\,\,\,\,\,
Y^{\prime \prime}_d = v_d \, \left(
\begin{array}{ccc}
a_d/v_d & 0 & 0 \\
0 & b_d & 0 \\
0 & 0 & c_d
\end{array}
\right)\,,
\end{equation}
where $v_{u,d}$ are vevs associated to the breaking of $D_{15}$, and $\lbrace a, b, c \rbrace_{u,d}$ are free ultraviolet ($D_{15}$-symmetric) Lagrangian parameters that can be directly related to the mass eigenvalues of the theory:  $a_u^2 \propto y_t^2$, $b_u^2 \propto y_u^2$, $c_u^2 \propto y_c^2$, and analogously for the down sector.  Plugging \eqref{eq:D15matrices} into \eqref{eq:upquark1}-\eqref{eq:delta}, we obtain predictions for the Yukawa eigenvalues and CKM mixing parameters that are equivalent to those from \cite{Bernigaud:2020wvn} (at particular numerical scan points) when appropriate mass hierarchies are realized, as expected.

As further confirmation, we have done a more exhaustive scan of \eqref{eq:leptobasis} without any information associated to the special $D_{15}$ encoded ---  $y_i^2$, $\theta_{ij}$, and $\delta$ are drawn randomly, as is the lone parametric degree of freedom in $\Lambda_d$ (we have assumed a special texture for the leptoquark coupling $y^{LL}$ in the fermion mass-eigenstate basis).  We again find exact agreement between the predictions of \eqref{eq:upquark1}-\eqref{eq:delta} and those values extracted numerically in the scans.

\section{\large{Renormalization Group Flow}}
\label{sec:RGE}

The formulae presented in Section \ref{sec:FORMULAE} have a number of foreseeable applications and benefits.  As a particularly powerful example, we now use \eqref{eq:upquark1}-\eqref{eq:delta} to perform an exact computation of the renormalization group flow (RGE) of Dirac flavour parameters in terms of invariants, thereby allowing these parameters to be computed at any scale in which the formulae in Section \ref{sec:FORMULAE} hold.

We begin from the top, and find the $\psi_{u}$ parameters' running is given by 
\begin{align}
\nonumber
 \psi_u^2 \, \dot{\psi}_u &= -2 I_1^2 \dot{I}_1 + 3(\dot{I}_1 \hat{I}_3 +I_1 \dot{\hat{I}}_3) - 3 \dot{\hat{I}}_6 \\
\label{eq:psirun}
& +  \frac{3\,\left(36 \hat{I}_3^2 \dot{\hat{I}}_3 + 18 \hat{I}_6 \dot{\hat{I}}_6 + 12 I_1^2 \dot{I}_1 \hat{I}_6 + 4 I_1^3 \dot{\hat{I}}_6  - 6 I_1 \dot{I}_1 \hat{I}_3^2 - 6 I_1^2 \hat{I}_3 \dot{\hat{I}}_3  - 18 \left(\dot{I}_1 \hat{I}_3 \hat{I}_6 + I_1 \dot{\hat{I}}_3 \hat{I}_6 + I_1 \hat{I}_3 \dot{\hat{I}}_6 \right) \right)}{2\,\left(\psi_u^3 + 2 I_1^3 - 9 I_1 \hat{I}_3 + 9 \hat{I}_6 \right)} \,,
\end{align}
and analogously for $\psi_d$ using \eqref{eq:downreplace}.  One notes that \eqref{eq:psirun} is given entirely in terms of the RGE of the flavour invariants themselves.  Here $\dot{x} \equiv \mu \,dx / d \mu$.  Using \eqref{eq:psirun}, one can also derive the RGE for the Yukawa eigenvalues, finding
\begin{align}
\nonumber
3\, \psi_u \, \dot{(y_i^2)} &=  -3 \,\dot{\psi}_u \, y_i^2  + (-2)^{1/3} \left(2 I_1 \dot{I}_1 -3 \dot{\hat{I}}_3 + (-2)^{-1/2}\left(\dot{I}_1 \psi_u + I_1 \dot{\psi}_u\right) + (-2)^{-2/3}2 \psi_u \dot{\psi}_u \right) \,, \\
\nonumber
12\, \psi_u \, \dot{(y_{j,k}^2)} &=  -12\,\dot{\psi}_u \,y_{j,k}^2  + 2 (-2)^{4/3} I_1 \dot{I}_1 - 3 (-2)^{4/3} \dot{\hat{I}}_3 + 4 (\dot{I}_1 \psi_u + I_1 \dot{\psi}_u) + 2(-2)^{2/3} \psi_u \dot{\psi}_u \\
\nonumber
&\mp  \frac{\dot{\psi}_u}{\psi_u} \,\left(24\, \psi_u^2 \left(I_1^2 - 3 \hat{I}_3 \right) + 6 \left(-2\right)^{5/3} \left(I_1^2 -3\, \hat{I}_3 \right)^2 -3  \left(-2\right)^{4/3} \psi _u^4\right)^{1/2} \\
\label{eq:y23run}
&\mp \psi_u^2 \frac{48\, I_1 \dot{I}_1 - 72\, \dot{\hat{I}}_3 - 6(-2)^{4/3} \psi_u \dot{\psi}_u +12(-2)^{5/3} \left(\frac{(I_1^2-3\hat{I}_3)(2I_1 \dot{I}_1 -3 \dot{\hat{I}}_3)}{\psi_u^2} - \frac{\dot{\psi}_u}{\psi_u^3} \left(I_1^2 -3 \hat{I}_3 \right)^2 \right)}{\left(96\, \psi_u^2 \left(I_1^2 - 3\, \hat{I}_3 \right) + 24 \left(-2\right)^{5/3} \left(I_1^2 -3\, \hat{I}_3 \right)^2 -12\left(-2\right)^{4/3} \psi _u^4\right)^{1/2} }
\end{align}
as well as the RGE for the mixing angles $s_{ij}$,
\begin{align}
\label{eq:s13run}
\dot{s}_{13} &= \frac{1}{2} \left[ -\frac{ \mathcal{P}_{13}}{s_{13}\, \Delta_{bd}^- \Delta_{bs}^- \Delta_{cu}^- \Delta_{ut}^-} - s_{13} \sum_{(ij)\, \in \, \mathbb{s}_1} \frac{\dot{\Delta}^-_{ij}}{\Delta^-_{ij}}\right] \,, \\
\label{eq:s23run}
\dot{s}_{23} &= \frac{1}{2}\left[\frac{\left(\Delta^{-}_{tu}  \mathcal{P}^a _{23} -s_{23}^2\, \Delta^{-}_{ct}\, \mathcal{P}^{b}_{23} \right)}{s_{23}\,\mathcal{D}_{23}} + s_{23}\left(\frac{\dot{\Delta}^{-}_{tu}}{\Delta^{-}_{tu}} - \frac{\dot{\Delta}^{-}_{ct}}{\Delta^{-}_{ct}} \right)\right]\,,\\
\label{eq:s12run}
\dot{s}_{12} &= \frac{1}{2}\left[ \frac{\mathcal{P}^a_{12}}{s_{12} \mathcal{D}_{12}} - s_{12} \left(\frac{\dot{\Delta}^{-}_{ds}}{\Delta^{-}_{ds}} + \Delta^{-}_{ds} \, \frac{\mathcal{P}^b_{12}}{\mathcal{D}_{12}} \right)\right]\,\,,
\end{align}
where, in the equation for $\dot{s}_{23}$,  the index set $\mathbb{s}_1$ is $\mathbb{s}_1 = \lbrace (bs), (bd), (cu), (ut) \rbrace$, and the denominator functions $\mathcal{D}_{ij}$ are given in \eqref{eq:s13}-\eqref{eq:s12}, defining $s_{ij}^2 \equiv \mathcal{N}_{ij}/\mathcal{D}_{ij}$. The remaining product functions $\mathcal{P}_{ij}$ appearing in \eqref{eq:s13run}-\eqref{eq:s12run} are tedious but algebraically simple expressions, and we give them in Appendix \ref{app:RGEFORM} for completeness.
Finally, one can derive a compact expression for the RGE of the Dirac phase $\delta$,
\begin{equation}
\label{eq:deltarun}
\dot{s}_\delta = s_\delta \left[ \frac{\dot{I}^{-}_{11}}{I_{11}^-} - \sum_{(ij)\, \in \, \mathbb{s}_2} \frac{\dot{\Delta}^-_{ij}}{\Delta^-_{ij}} - \dot{s}_{12} \frac{(1-2 \,s_{12}^2)}{s_{12}c_{12}^2} - \dot{s}_{23} \frac{(1-2\, s_{23}^2)}{s_{23}c_{23}^2}- \dot{s}_{13} \frac{(1-3 \,s_{13}^2)}{s_{13}c_{13}^2}\right]\,,
\end{equation}
which depends on \eqref{eq:s12run}. Here the index set $\mathbb{s}_2$ is over the mass-differences appearing on the RHS of \eqref{eq:delta}, $\mathbb{s}_2 = \lbrace (tc), (tu), (cu), (bs), (bd), (sd) \rbrace$.

\begin{figure}[tp]
\centering
\includegraphics[width=80mm]{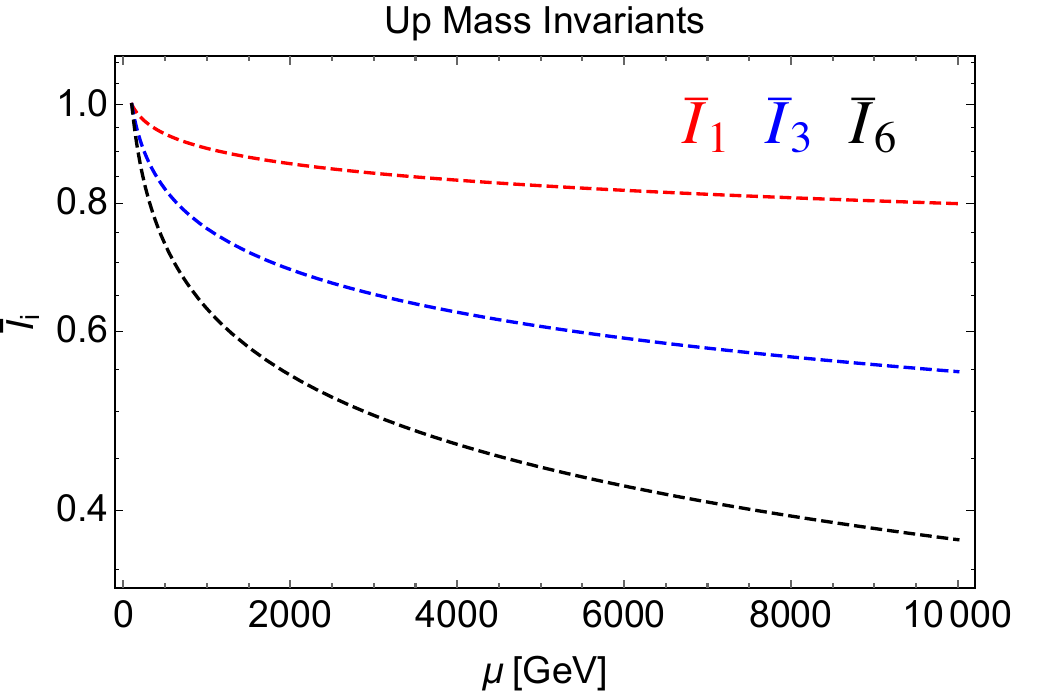}
\includegraphics[width=80mm]{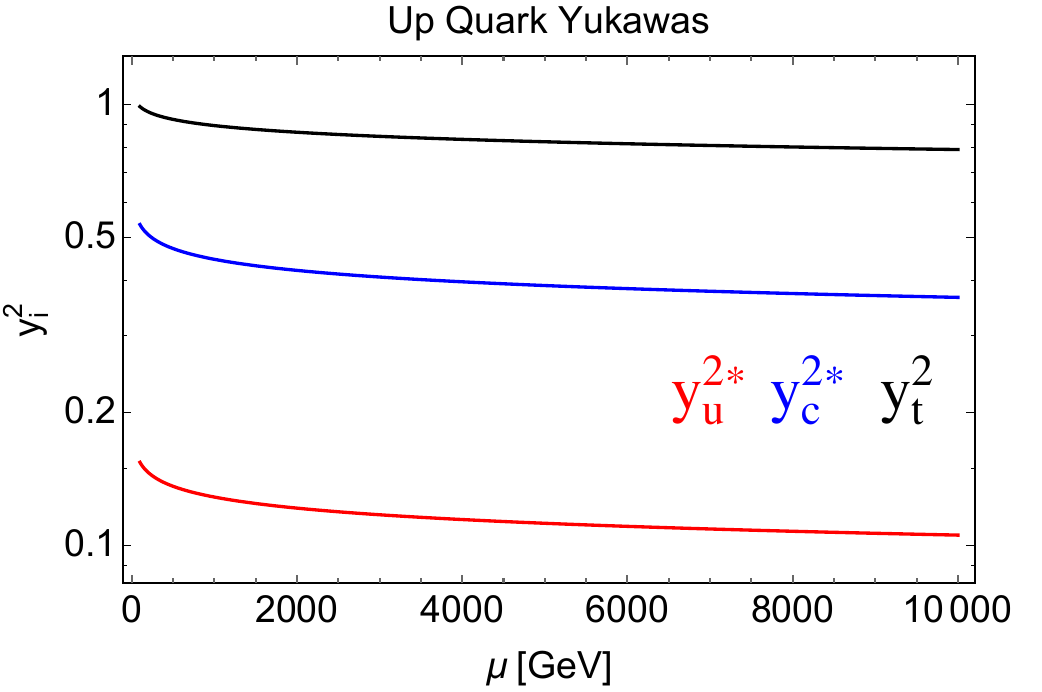} \\
\includegraphics[width=80mm]{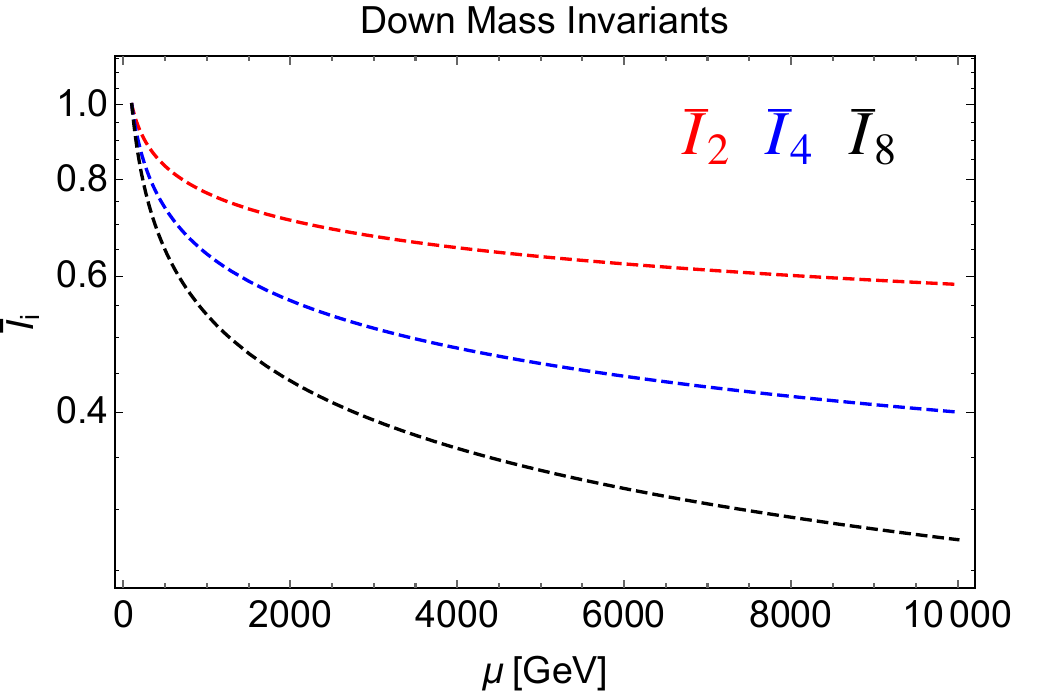}
\includegraphics[width=80mm]{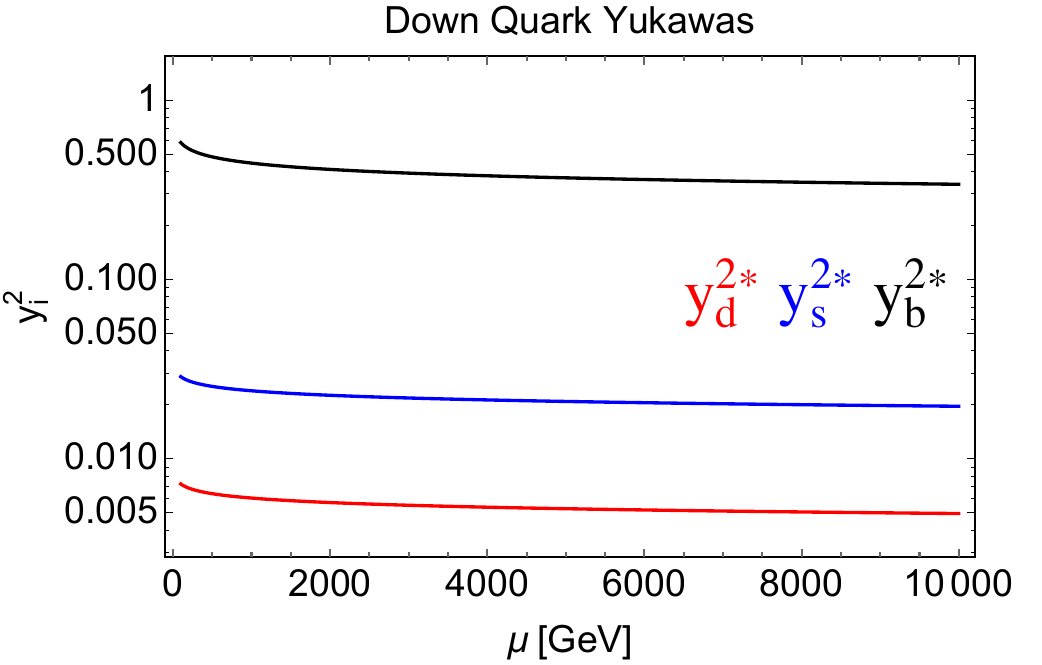}
\caption{TOP LEFT: Running of the (normalized) unmixed invariants relevant for up-quark Yukawa extraction at one-loop order. TOP RIGHT:  The same for the actual up-quark Yukawa eigenvalues (squared).  Note that we have rescaled the up and charm eigenvalues by $10^9$ and $10^4$ respectively, hence the starred ($\star$) notation.  BOTTOM LEFT:  Running of the (normalized) unmixed invariants relevant for down-quark Yukawa extraction at one-loop order.    BOTTOM RIGHT:  The same for the actual down-quark Yukawa eigenvalues (squared).  We have again rescaled the eigenvalues for the purpose of plotting, multiplying the bottom, strange, and down eigenvalues by $10^3$, $10^5$ and and $10^7$ respectively.}
\label{fig:MassRun}
\end{figure}

\begin{figure}[tp]
\centering
\includegraphics[width=80mm]{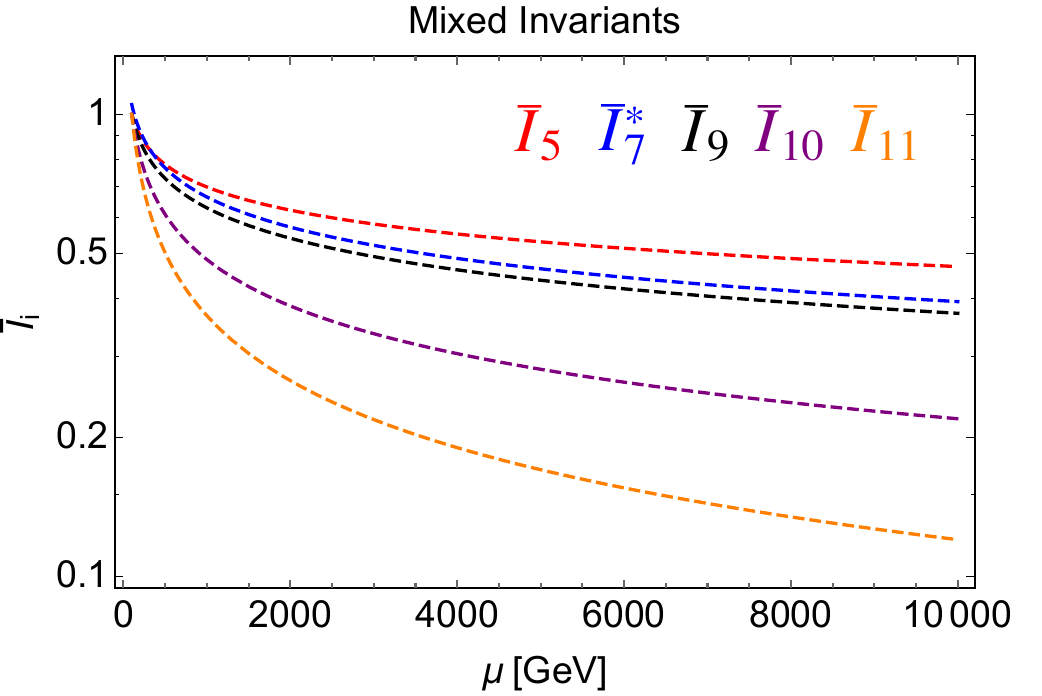}
\includegraphics[width=80mm]{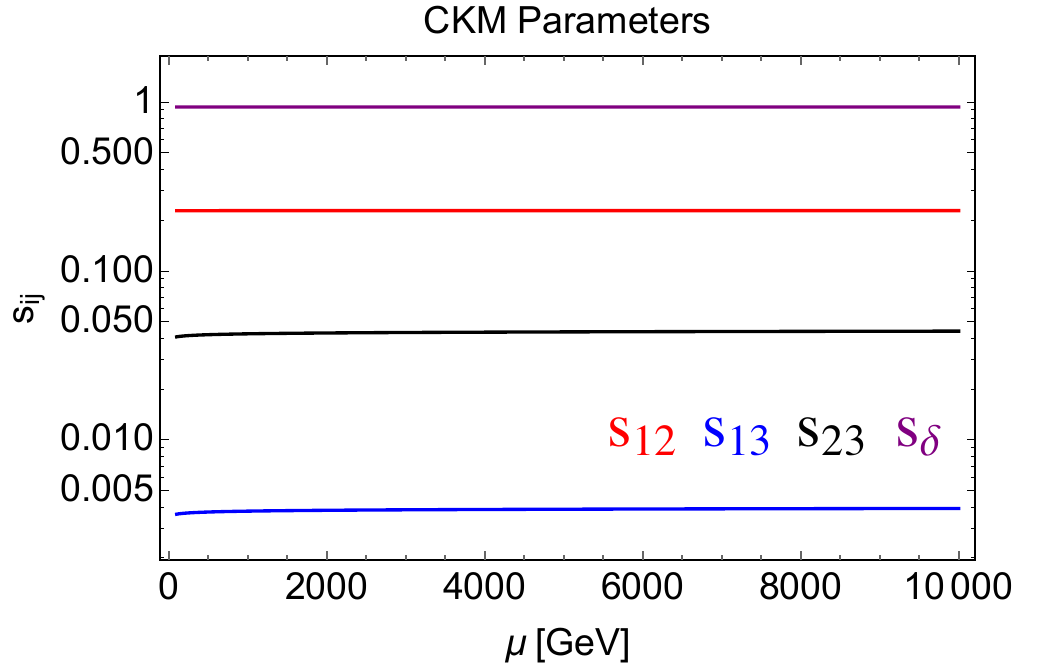}
\caption{LEFT: Running of the (normalized) mixed invariants relevant for extracting the CKM mixing elements, at one-loop order. Note that $\bar{I}_7$ has been multiplied by 1.05 in order to visually distinguish it from $\bar{I}_5$.  RIGHT:  Running of the mixing angles and Dirac CP-violating phase of the CKM matrix, again at one-loop order}
\label{fig:MixRun}
\end{figure}

The RGE derived in \eqref{eq:psirun}-\eqref{eq:deltarun} are again exact, and valid to all orders in the $\overline{v}_T / \Lambda$ geoSMEFT expansion, and to all orders in the perturbative loop expansion (which is implicitly embedded in $\dot{I}_i$).  They therefore open the door to any number of novel phenomenological applications that depend on the radiative structure of flavour parameters in a (geo)SMEFT context.  Furthermore, one sees that, as must be the case given \eqref{eq:upquark1}-\eqref{eq:delta}, the running of all flavour parameters is given entirely by the running of the flavour invariants $I$. Indeed, it is within the specific forms of $\dot{I}$ that one would find contributions due to operator mixing induced from objects outside of the Yukawa sector (e.g. from a chromo-dipole term $\mathcal{O} \sim \overline{\psi}_p \sigma^{\mu \nu} t^a \psi_r G^a_{\mu\nu}$), that might alter \eqref{eq:HFops} upon RGE.

\subsection{Computation of Quark Sector RGE at One-Loop Order}
\label{sec:LOOP}

Since the RGE for each invariant was derived explicitly in \cite{Feldmann:2015nia} in the quark sector (assuming MFV particle spectrum --- see Appendix \ref{app:RGEFORM}), it is now also straightforward to compose the system of coupled differential equations, and obtain exact (numerical) solutions for the running of quark mass eigenvalues, mixing angles, and CP-violating phase at arbitrary mass-dimension orders in the (geo)SM(EFT).  We have done so using SM couplings, and show the results in Figure \ref{fig:MassRun} for the unmixed invariants and associated Yukawa eigenvalues.  Note that we have set the boundary conditions at $\mu = 100$ GeV for these RGE using the values reported for the mass and mixing parameters in the 2020 PDG databook \cite{Zyla:2020zbs}, and have neglected the electroweak correction (as in  \cite{Feldmann:2015nia}) and the running of the strong coupling for this illustration.  For the invariants we have normalized all of the functions to their values at $\mu = 100$ GeV, i.e. $\bar{I}_i = \hat{I}_i / \hat{I}_i(100)$ and similarly for the un-hatted invariants in \eqref{eq:massinvariants}-\eqref{eq:CKMinvariants}.  Then, in Figure \ref{fig:MixRun} we show the running of the mixed variants relevant to extracting information about the CKM mixing matrix (left panel), and then the CKM mixing elements themselves (right panel).  Note that while we have used the one-loop RGE in the above example calculation, two-loop expressions for $\dot{I}_{i}$ are also provided in \cite{Feldmann:2015nia}, and hence performing the analogous computation of the flavour parameter evolution is no more difficult.


\section{\large{Summary and Outlook}}
\label{sec:CONCLUDE}
We have reported a set of compact, exact formulae for computing Dirac Yukawa/mass eigenvalues and CKM mixing parameters directly from $3 \times 3$ Yukawa couplings in arbitrary flavour bases.  These formulae hold at all orders in the $\overline{v}_T / \Lambda$ expansion characterizing the geoSMEFT, and therefore complete the list of geometric expressions originally given in \cite{Helset:2020yio} in the Dirac flavour sector.  In addition to demonstrating their utility in (e.g.) BSM model-building, we have also used our expressions to derive analytic RGE for Dirac masses and mixings at all orders in $\overline{v}_T / \Lambda$, and all perturbative loop orders.  The practical implementation of these latter formulae requires knowledge of the RGE of the 11 flavour invariants that exclusively compose them, which has been given at one- and two-loop perturbative orders for quark flavour invariants in MFV theories \cite{Feldmann:2015nia}.  We use the results from \cite{Feldmann:2015nia} to perform an explicit one-loop numerical computation within our formalism.

The results presented in this paper open the door to any number of interesting physics applications, from aiding global SMEFT fits to experimental data in the flavour sector (cf. \cite{Descotes-Genon:2018foz,Bruggisser:2021duo}), to studying non-standard RGE-induced flavour violation effects in the SMEFT (cf. \cite{Aebischer:2020lsx}), to permitting exact predictions in flavoured model building (cf. Section \ref{sec:BSM}).  In addition to these foreseeable applications, we also note that an extension of our formalism to the lepton sector would be desirable for describing (e.g.) neutrino mass and mixing to all orders in $\overline{v}_T / \Lambda$ with the geoSMEFT.  Of course, it is clear that describing the mass sector of the charged leptons at all orders is already achieved through \eqref{eq:massinvariants}, by replacing $\mathbb{Y}_{u,d} \rightarrow \mathbb{Y}_e$.  Furthermore, by $(a)$ adding to the SM a neutrino Yukawa operator of the form $\mathcal{L} \supset Y_N \, L \tilde{H}^\dagger \overline{N}$, with $N$ a singlet neutrino transforming under $\mathcal{G}_{SM}$ as $N \sim \left({\bf{1}}, {\bf{1}}, 0\right)$, and $(b)$ forbidding lepton-number violation, one can also use our formalism to describe leptonic mixing, for renormalizable interactions only.  To obtain an all-orders description in this case one must first develop a geo$\nu$SMEFT.  On the other hand, a formal treatment of Majorana neutrinos is in principle achievable with the geoSMEFT alone, if an all-orders field connection can be formalized for a tower of operators whose lowest-dimension contribution is the dim-5 Weinberg operator \cite{Weinberg:1979sa}.  Then leptonic mixing will be described by six parameters, accounting for the two additional Majorana phases.  A set of 15 polynomially independent invariants, as required in this case, has already been given in \cite{Jenkins:2009dy}.  Upon writing down the hypothetical `Weinberg connection,' one must then find an amenable basis for deriving relations analogous to \eqref{eq:massinvariants}-\eqref{eq:phaseinvariants}, which has recently been studied outside of a (geo)SMEFT context in \cite{Wang:2021wdq}.  Work on deriving the RGE of leptonic flavour invariants can also be found in \cite{Wang:2021wdq,Yu:2020gre}.  We leave pursuing such interesting extensions in a geoSMEFT context to future work.

\section*{\large{Acknowledgements}}
JT and MT are supported by the Villum Fund, project number 00010102.  We thank Andreas Helset for helpful insight early in the project, and JT acknowledges interesting discussions with Fady Bishara, Tyler Corbett, Christophe Grojean, Wolfgang G. Hollik, and Ayan Paul on related topics.
\begin{appendix}
\section{\large{Formulae for RGE Computation}}
\label{app:RGEFORM}
In this appendix we report some of the functions employed in Section \ref{sec:RGE} for completeness.
\subsection{All-Orders Product Functions}
In the derivation of the all-orders expressions for the mixing angles' RGE, cf. \eqref{eq:s13run}-\eqref{eq:s12run}, product functions appear.  They are given by
\begin{align}
\nonumber
\mathcal{P}_{13} &=
\nonumber
y_u^2 \left(\dot{\hat{I}}_9 + \dot{(y_b^2)} \left(\hat{I}_5 - y_b^2 \Delta_{ct}^+ \right) + y_b^2 \left(\dot{\hat{I}}_5 - \dot{(y_b^2)}\, \Delta^+_{ct} - y_b^2 \dot{\Delta}^+_{ct} \right) - \dot{(y_d^2)}\,y_s^2 \Delta^+_{ct} - y_d^2\, \dot{(y_s^2)} \,\Delta^+_{ct} - y_d^2 y_s^2 \dot{\Delta}^+_{ct} \right) \\
 \nonumber
&+\dot{\hat{I}}_{10} + \dot{(y_b^2)}\left( \hat{I}_7 - \Delta^+_{ds} \Delta^+_{uc} \Delta^+_{ut}\right) + \dot{(y_u^2)} \left(\hat{I}_9 + y_b^2 \left(\hat{I}_5 - y_b^2 \Delta^+_{ct} \right) -y_d^2 y_s^2 \Delta^+_{ct}\right)\\
\nonumber
&+ y_b^2 \left(\dot{\hat{I}}_7 - \dot{\Delta}^+_{ds}\Delta^+_{uc}\Delta^+_{ut} -\Delta^+_{ds}\dot{\Delta}^+_{uc}\Delta^+_{ut}-\Delta^+_{ds}\Delta^+_{uc}\dot{\Delta}^+_{ut}  \right) \,,
 \end{align}
 for the (13) angle,
 \begin{align}
 \nonumber
 \mathcal{P}_{23}^a &=
 \nonumber
 y_b^2 \left( \dot{(y_c^2)} \left(\Delta_{ct}^+ \Delta_{ds}^+ -\hat{I}_5\right) +  y_c^2 \left(\dot{\Delta}_{ct}^+ \Delta_{ds}^+ + \Delta_{ct}^+ \dot{\Delta}_{ds}^+ - \dot{\hat{I}}_5 \right) + \dot{(y_u^2)} \Delta_{ct}^+ \Delta_{ds}^+ +  y_u^2  \dot{\Delta}_{ct}^+ \Delta_{ds}^+ +  y_u^2 \Delta_{ct}^+ \dot{\Delta}_{ds}^+ - \dot{\hat{I}}_7\right) \\
 \nonumber
 &-\dot{\hat{I}}_{10} + \dot{(y_c^2)} \left(\Delta_{ut}^+ \left(y_b^4 + y_d^2 y_s^2 \right) - \hat{I}_9 \right) +  \dot{(y_b^2)} \left(y_c^2 \left(\Delta_{ct}^+ \Delta_{ds}^+ - \hat{I}_5 \right) + y_u^2 \Delta_{ct}^+ \Delta_{ds}^+ - \hat{I}_7 \right)  \\
 \nonumber
 &+ y_c^2 \left(\dot{\Delta}_{ut}^+ \left(y_b^4 +y_d^2 y_s^2\right) + \Delta_{ut}^+ \left(\dot{(y_b^4)} + \dot{(y_d^2)}y_s^2 + y_d^2 \dot{(y_s^2)} \right) - \dot{\hat{I}}_9 \right) \,,\\
 \nonumber
 \mathcal{P}_{23}^b &= \dot{\hat{I}}_{10} + \dot{(y_u^2)} \hat{I}_9 + y_u^2 \dot{\hat{I}}_9-\left(\dot{(y_u^4)} +\dot{(y_c^2)} y_t^2 + \dot{(y_t^2)}y_c^2 \right) \left(y_b^4+y_d^2 y_s^2 \right) - \left(\dot{(y_b^4)} +\dot{(y_d^2)} y_s^2 + \dot{(y_s^2)}y_d^2 \right) \left( y_u^4+y_c^2 y_t^2\right) \\
 \nonumber
 &+ \dot{(y_b^2)}\left(\hat{I}_7 + y_u^2 \left(\hat{I}_5 - 2 \Delta_{ct}^+ \Delta_{ds}^+ \right) \right) + y_b^2 \left(\dot{\hat{I}}_7 + \dot{(y_u^2)} \left(\hat{I}_5 - 2 \Delta_{ct}^+ \Delta_{ds}^+ \right) + y_u^2 \left(\dot{\hat{I}}_5 - 2 \dot{\Delta}_{ct}^+ \Delta_{ds}^+ -2 \Delta_{ct}^+ \dot{\Delta}_{ds}^+ \right) \right) \,,
\end{align}
for the (23) angle, and finally
\begin{align}
\nonumber
\mathcal{P}_{12}^a  &= \dot{\Delta}_{db}^- \left(\hat{I}_{10} + y_s^2 \left(\hat{I}_7 - y_c^2 y_t^2 \Delta_{db}^+ \right) \right) +  \dot{(y_u^2)} \Delta_{bd}^- \left(\Delta_{sb}^+ \Delta_{ct}^+ \Delta_{ds}^+ - y_s^2 \hat{I}_5 - \hat{I}_9 \right) \\
\nonumber
&+  \Delta_{db}^- \left(\dot{\hat{I}}_{10} + \dot{(y_s^2)} \left(\hat{I}_7 - y_c^2 y_t^2 \Delta_{db}^+ \right) + y_s^2 \left(\dot{\hat{I}}_7 - \dot{(y_c^2)} y_t^2 \Delta_{db}^+ - y_c^2 \dot{(y_t^2)} \Delta_{db}^+ - y_c^2 y_t^2 \dot{\Delta}_{db}^+  \right) \right) \\
\nonumber
&+y_u^2 \Delta_{bd}^- \left(-\dot{\hat{I}}_9 - \dot{(y_s^2)} \hat{I}_5 - y_s^2 \dot{\hat{I}}_5 + \dot{\Delta}_{sb}^+ \Delta_{ct}^+ \Delta_{ds}^+ + \Delta_{sb}^+ \dot{\Delta}_{ct}^+ \Delta_{ds}^+ + \Delta_{sb}^+ \Delta_{ct}^+ \dot{\Delta}_{ds}^+\right) \\
\nonumber
&+\dot{(y_u^4)} y_s^2 \left(y_b^4 - y_d^4 \right) +y_u^4 \dot{(y_s^2)} \left(y_b^4 - y_d^4 \right) +y_u^4 y_s^2 \left(\dot{(y_b^4)} - \dot{(y_d^4)} \right) + y_u^2 \dot{\Delta}_{bd}^- \left(\Delta_{sb}^+ \Delta_{ct}^+ \Delta_{ds}^+ - y_s^2 \hat{I}_5 - \hat{I}_9 \right)\,, \\
\nonumber
\mathcal{P}_{12}^b &=
\nonumber
-\left(\dot{(y_b^4)} +\dot{(y_d^2)} y_s^2 + \dot{(y_s^2)}y_d^2 \right) \left(y_u^4+y_c^2 y_t^2 \right) - \left(\dot{(y_u^4)} +\dot{(y_c^2)} y_t^2 + \dot{(y_t^2)}y_c^2 \right) \left( y_b^4+y_s^2 y_d^2\right) \\
\nonumber
&+y_b^2 \left( \dot{\hat{I}}_7 + \dot{(y_u^2)} \left(\hat{I}_5 -2 \Delta_{ct}^+ \Delta_{ds}^+ \right) + y_u^2 \left(\dot{\hat{I}}_5 -2 \Dot{\Delta}_{ct}^+ \Delta_{ds}^+ -2 \dot{\Delta}_{ds}^+ \Delta_{ct}^+ \right)\right)  \\
\nonumber
&+ \dot{\hat{I}}_{10} + \dot{(y_u^2)} \hat{I}_9 + y_u^2 \dot{\hat{I}}_9 + \dot{(y_b^2)} \left(\hat{I}_7 + y_u^2 \left(\hat{I}_5 -2 \Delta_{ct}^+ \Delta_{ds}^+ \right) \right)
\end{align}
for the (12) angle.  While somewhat tedious, these expressions amount to simple polynomials in the Yukawa eigenvalues and flavour invariants, and it is clear that every individual term comes with only a single derivative operator.
Note also that we have been careful to indicate when the derivatives are of the \emph{squared} $\dot{\left(y_i^2\right)}$ or \emph{quartic} $\dot{\left(y_i^4\right)} = 2 y_i^2 \dot{\left(y_i^2\right)}$ Yukawa eigenvalues, and it is of course clear that $\dot{\Delta}^{\pm}_{ij} = \dot{(y_{i}^2)} \pm \dot{(y_{j}^2)}$.
\subsection{One-Loop RGE for $I_{i}$}
The RGE equations for $\hat{I}_{i}$ were derived, for MFV theories, in \cite{Feldmann:2015nia} at one- and two-loop perturbative order.  We have used the one-loop expressions in the sample numerical calculation performed in Section \ref{sec:LOOP}, and so we report the results from \cite{Feldmann:2015nia}:
\begin{align}
\nonumber
\mu \frac{d I_1}{d\mu} & \simeq 2 a_0 I_1 + 2 a_1 \left(\frac{2}{3} I_1^2 -2 \hat{I}_3\right) + 2 a_2 \left(\hat{I}_5 - \frac{I_1 I_2}{3} \right)\,,\\
\nonumber
\mu \frac{d I_2}{d \mu} & \simeq 2 b_0 I_2 + 2 b_1 \left(\frac{2}{3} I_2^2 -2 \hat{I}_4\right) + 2 b_2 \left(\hat{I}_5 - \frac{I_1 I_2}{3} \right) \,,\\
\nonumber
\mu \frac{d \hat{I}_3}{d \mu} & \simeq 4 a_0 \hat{I}_3 -2 a_1 \left(3 \hat{I}_6 - \frac{I_1 \hat{I}_3}{3} \right) -2 a_2 \left(\hat{I}_7 -\frac{I_2 \hat{I}_3}{3} \right) \,,\\
\nonumber
\mu \frac{d \hat{I}_4}{d \mu} & \simeq 4 b_0 \hat{I}_4 -2 b_1 \left(3 \hat{I}_8 - \frac{I_2 \hat{I}_4}{3} \right) -2 b_2 \left(\hat{I}_9 -\frac{I_1 \hat{I}_4}{3} \right) \,,\\
\nonumber
\mu \frac{d \hat{I}_5}{d \mu} & \simeq \left(2 a_0 + 2 b_0 \right)\hat{I}_5 + \left(2 a_1 + 2 b_2 \right) \left(\hat{I}_7 - I_2 \hat{I}_3 + \frac{2 I_1 \hat{I}_5}{3} \right) + \left(2 a_2 + 2 b_1 \right) \left(\hat{I}_9 - I_1 \hat{I}_4 + \frac{2 I_2 \hat{I}_5}{3} \right) \,,\\
\nonumber
\mu \frac{d \hat{I}_6}{d \mu} & \simeq 6 a_0 \hat{I}_6 \,,\\
\nonumber
\mu \frac{d \hat{I}_7}{d \mu} & \simeq \left(4 a_0 +2 b_0 \right) \hat{I}_7 + \left(2 a_1 - 2 b_2 \right)\left(\frac{I_1 \hat{I}_7}{3} - I_2 \hat{I}_6\right) - \left(2a_2 - 2 b_1\right) \left(\hat{I}_{10} - \hat{I}_3 \hat{I}_4 + 2\frac{I_2 \hat{I}_7}{3} \right) \,, \\
\nonumber
\mu \frac{d \hat{I}_8}{d \mu} & \simeq 6 b_0 \hat{I}_8 \,, \\
\nonumber
\mu \frac{d \hat{I}_9}{d \mu} & \simeq \left(4 b_0 +2 a_0 \right) \hat{I}_9 + \left(2 b_1 - 2 a_2 \right)\left(\frac{I_2 \hat{I}_9}{3} - I_1 \hat{I}_8\right) - \left(2b_2 - 2 a_1\right) \left(\hat{I}_{10} - \hat{I}_3 \hat{I}_4 + 2\frac{I_1 \hat{I}_9}{3} \right) \,,\\
\nonumber
\mu \frac{d \hat{I}_{10}}{d \mu} & \simeq \left(4 a_0 + 4 b_0 \right) \hat{I}_{10} + \left(2 a_1 + 2 b_2 \right)\left(\frac{I_1 \hat{I}_{10}}{3} - \hat{I}_4 \hat{I}_6\right) + \left(2a_2 + 2 b_1\right) \left(\frac{I_2 \hat{I}_{10}}{3} - \hat{I}_3 \hat{I}_8\right) \,,\\
\label{eq:RGE1loop}
\mu \frac{d I^-_{11}}{d \mu} & \simeq \left(6 a_0 +6 b_0 + 2 a_1 I_1 + 2 b_1 I_2\right) I_{11}^- \,,
\end{align}
where the $a$ and $b$ parameters are defined as
\begin{align}
\nonumber
a_0 &= \frac{3}{8 \pi^2} \left(I_1 + I_2 + \frac{I_1-I_2}{2 n_g} \right) - 2 \frac{\alpha_s}{\pi}, \,\,\,\,\,\,\,\,\,\,\,\,\,\,\,\, a_1 = \frac{3}{16\pi^2} \,, \,\,\,\,\,\,\,\,\,\,\,\,\,\,\, a_2 = -\frac{3}{16\pi^2} \,, \\
\label{eq:abparameters}
b_0 &= \frac{3}{8 \pi^2} \left(I_1 + I_2 + \frac{I_2-I_1}{2 n_g} \right) - 2 \frac{\alpha_s}{\pi}, \,\,\,\,\,\,\,\,\,\,\,\,\,\,\,\, b_1 = \frac{3}{16\pi^2} \,, \,\,\,\,\,\,\,\,\,\,\,\,\,\,\, b_2 = -\frac{3}{16\pi^2} \,,
\end{align}
which depend on the strong coupling constant $\alpha_s$ and the number of generations $n_g$.
Note that the expressions in \eqref{eq:RGE1loop} - \eqref{eq:abparameters} neglect subdominant electroweak corrections.

\end{appendix}


\end{document}